# Internal mobility edge in doped graphene: frustration in a renormalized lattice

Gerardo G. Naumis
*Departamento de Física-Química, Instituto de Física,*
*Universidad Nacional Autónoma de México (UNAM) Apartado Postal 20-364, 01000, México D.F., MEXICO.*
(Dated: July 4, 2007)

We show that an internal localization mobility edge can appear around the Fermi energy in graphene by introducing impurities in the split-band regimen, or by producing vacancies in the lattice. The edge appears at the center of the spectrum and not at the band edges, in contrast with the usual picture of localization. Such result is explained by showing that the bipartite nature of lattice allows to renormalize the Hamiltonian, and the internal edge appears because of frustration effects in the renormalized lattice. The size in energy of the spectral region with localized states is similar in value to that observed in narrow gap semiconductors.

PACS numbers: 81.05.Uw, 73.22.-f, 73.21.-b,

Only very recently a two dimensional form of carbon was obtained [1]. This material, known as graphene has attracted a lot of research due to its amazing electrical and mechanical properties [2]. For example, electrons in graphene behave as massless relativistic fermions that satisfy the Dirac equation [4]. Such property is a consequence of the bipartite crystal structure [5], in which a linear dispersion relationship appears at the center of the electronic spectrum. Also, one can cite the high mobility of its charge carriers that remains higher even at high electric-field induced concentration, that translates into ballistic transport on a submicron scale [3] at $300°K$. These and other unusual electronic properties of graphene makes it a promising material for building electronic devices. However, from the point of view of applications, the use of pure graphene pose some problems. The transmission probability of electrons across a potential barrier is always unity, irrespective of the height and width of the barrier. This behavior is related to the Klein paradox in relativistic quantum mechanics [2]. As a result, conductivity can not be changed by an external gate voltage, a feature required to build a FET transistor, although a quantum dot has been used to perform the required task [6]. In spite of all this research in pure graphene, at the moment there is not so much knowledge in the effects of impurities in the electronic properties and its potential use to produce gates. In a previous work [7], the density of states (DOS) of graphene with Anderson type of disorder revealed that the linear dispersion relationship was affected [7], and recently many electrical properties of graphene with disorder have been obtained [8]. However, the existance of a mobility edge has not adressed. Here we show that graphene doped with impurities or with vacancies presents a very unusual property; instead of having a localization mobility edge at the band limits as in the usual Anderson localization, the localized states appear at the center of the spectrum, around the Fermi energy. As we will show, this is a simple consequence of the bipartite crystal structure, which produces a frustration effect in a renormalized Hamiltonian that removes one of the bipartite lattices. The observed effect can be used in certain applications, since the mobility edge can be tuned with a given concentration of impurities.

Let us start by considering the tight-binding Hamiltonian of graphene with disorder, which can be written as $H = H_0 + H_1$, where $H_0$ is the pure graphene Hamiltonian given by [9],

$$H_0 = E_0 \sum_i |i\rangle\langle i| + \gamma_0 \sum_{<i,j>} |i\rangle\langle j| + H_1. \quad (1)$$

$E_0$ is the self-energy of carbon and $\gamma_0$ is the carbon-carbon resonance integral, as given in Ref. [9]. $H_1$ is the Hamiltonian of the perturbation due to defects,

$$H_1 = \delta E \sum_i |i\rangle\langle i| + \delta\gamma_0 \sum_{<i,j>} |i\rangle\langle j|, \quad (2)$$

where we define $\delta E \equiv E_I - E_0$ and $\delta\gamma_0 \equiv \gamma_I - \gamma_0$. Here $E_I$ is the self-energy of the defects, and $\gamma_I$ the transfer integral between impurities (which are basically isolated in the dilute limit). When $\delta E >> E_0$, the spectrum is divided in two parts, one centered around $E_0$ and the other at $E_0 + \delta E$. This case is known as the split-band limit. The states in the sub-band around the carbon self energy $E_0$, that we call the $C$-band, are strongly confined on carbon atoms. Furthermore, in the limit $\delta E >> E_0$, it has been shown that impurity atoms can be formally removed in a tight-binding Hamiltonian [10], and thus the $C$-band can be studied by using a Hamiltonian restricted to $C$ sites only,

$$H_{CC} = E_0 \sum_{i \in C} |i\rangle\langle i| + \gamma_0 \sum_{i,j \in C} |i\rangle\langle j|. \qquad (3)$$

This Hamiltonian describes an electron that can hop from one site to its neighbors only if both are carbon atoms ($C$). Furthermore, the problem for the $C$ sub-band is similar to a lattice with holes, because impurity atoms act as perfect barriers in the limit of infinite self-energy. As a result, the results presented here are also valid for vacancies in the lattice.

Now let us study the spectrum of $H_{CC}$. First it is convenient to work on a renormalized Hamiltonian $H_{CC}$, which takes advantage of the bipartite nature of the $C$ lattice, once the $I$ atoms are removed. The bipartite character of the $C$ lattice means that it can be separated in two inter-penetrating sublattices, $A$ and $B$. We define two orthogonal operators that project the wavefunctions into each sublattice,

$$P_A = \sum_{i \in A} |i\rangle\langle i|, \text{ and } P_B = \sum_{j \in B} |j\rangle\langle j| \qquad (4)$$

Therefore, any eigenvector $|\phi\rangle$ of $H_{CC}$ can be written in terms of these projectors,

$$H_{CC}(P_A + P_B)|\phi\rangle = E(P_A + P_B)|\phi\rangle. \qquad (5)$$

Since $H_{CC}$ produces a hopping in the wave-function between the $A$ and $B$ sublattices,

$$H_{CC} P_A |\phi\rangle = E P_B |\phi\rangle, \text{ and } H_{CC} P_B |\phi\rangle = E P_A |\phi\rangle. \qquad (6)$$

From these equations, one can see that the spectrum is symmetric around $E = E_0$, since if $(P_A + P_B)|\phi\rangle$ is an eigenvector with eigenvalue $E$, $(P_A - P_B)|\phi\rangle$ is also an eigenvector with eigenvalue $-E$. We can decuple the sublattices by further applying $H_{CC}$ to Eqs. (6),

$$H_{CC}(H_{CC}(P_i|\phi\rangle)) = H_{CC}^2(P_i|\phi\rangle) = E^2(P_i|\phi\rangle), \qquad (7)$$

where $i = A$, $B$. Thus, the projection of an eigenvector in each sublattice is a solution of the squared Hamiltonian. Observe that the eigenvalues of $H_{CC}^2$ are positive definite, and their eigenstates are, at least, doubly degenerate. This spectrum can be regarded as the folding of the original spectrum of $H_{CC}$ around $E = 0$, in such a way that the two band edges of $H_{CC}$, are mapped into the highest eigenvalue of $H_{CC}^2$, while the states at the center of the original band are now at the minimum eigenvalue of the squared Hamiltonian ($E^2$). The important property of the renormalized Hamiltonian $H_{CC}^2$ is that the states at the bottom of the spectrum have an antibonding nature (the phase between neighbors is $\pi$), and we can expect that the frustration of the wavefunction can prevent the spectrum from reaching its minimum eigenvalue in a continuous form when frustration is present [11][12]. In fact, frustration acts as an effective potential which leads to localization since the wave-function tends to avoid regions of higher-frustration. The mobility edge appears when the energy cost in localization is less than that of having amplitude in frustrated bonds. As we will show next, this frustration augments with disorder. To do this, we observe that the Hamiltonian $H_{CC}^2$ is equivalent to a renormalization of sites $B$ in the lattice, which leads to a triangular lattice with an effective interaction, as shown in Fig. 1a). The new lattice contains odd rings, and when impurities are present, there are holes, as indicated in Fig. 1b). The corresponding Schröedinger equation derived from $H_{CC}^2$ is,

$$\left((E - E_0)^2 - Z_i \gamma_0^2\right) c_i(E) = \gamma_0^2 \sum_{j,i) \epsilon A} c_j(E), \qquad (8)$$

where $c_i(E)$ is the amplitude of the wave-function at site $i$ for an eigenenergy $E$, and the notation $(j,i)\epsilon A$ means that the sum is taken only for carbon atoms which are first neighbours in the new triangular lattice, i.e., those carbon atoms that were second neighbours in the original lattice. Notice that such atoms belong to only one of the bipartite sublattices $A$ or $B$. Due to the symmetry of the problem, we can solve for any sublattice, say for example sublattice $A$. Finally, $Z_i$ is the coordination number at site $i$. This number goes from 0 when a carbon atom is surrounded by impurities, to 3 in the lattice without defects. Then we can perform a variational procedure to estimate the ground state of Eq. (8). After multiplying Eq. (8) by $c_i^*(E)$ and summing over $i$,

$$(E - E_0)^2 = \sum_i Z_i \gamma_0^2 |c_i(E)|^2 + \gamma_0^2 \sum_i \sum_{j,i)\epsilon A} c_i^*(E) c_j(E), \qquad (9)$$

The first contribution is an effective self-energy while the second depends on the number of bonds and the amplitude and phase of the wave-function. For example, in pure carbon $Z_i = 3$. Also, the lattice is periodic from where we can write $c_j(E) = c e^{i\phi_j}$ where $c$ is an amplitude (in fact $c = 1/\sqrt{N}$ where $N$ is the number of atoms), and $\phi_j$ is a phase. The minimal eigenvalue is thus obtained from Eq. (8) when the phase difference between sites in the $A$ sublattice is $\pi$. Thus, the ground state has an antibonding nature and $c_i^*(E)c_j(E) = -1/N$. Using that there are three second neighbours for each atom, it follows that $E = E_0$ . As a consequence, this shows that there is no gap for pure graphene, as expected. However, the previous case reveals an interesting fact, the zero gap is obtained due to the balance between the positive renormalized self-energy $Z_i$ and the antibonding contribution. In pure graphene, both contributions match exactly to produce a gapless spectrum.

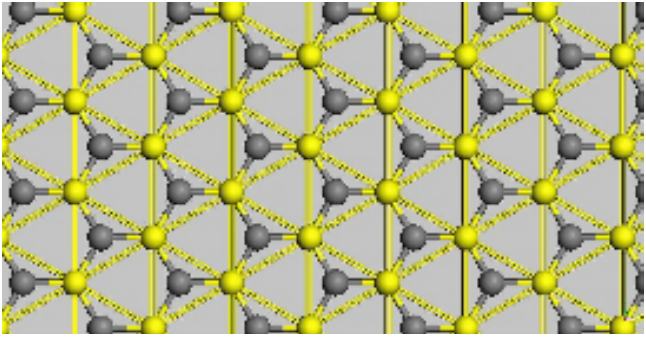

FIG. 1: Renormalization of the graphene lattice. Atoms in the $A$ sublattice are shown with different color than those in the $B$ sublattice . The new lattice that appears after renormalizing the $B$, is represented with double bonds.

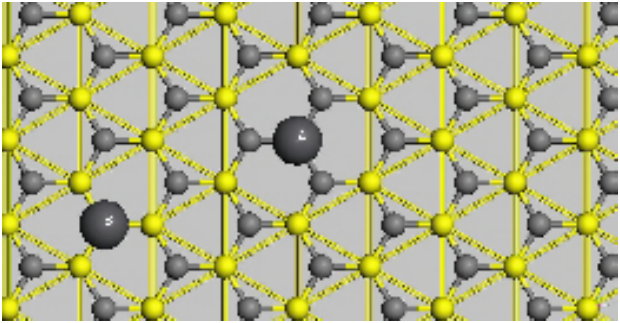

FIG. 2: Renormalization of the lattice with defects. The impurities are shown with dark color. There are two cases: the impurities can fall in the $A$ sublattice or in the $B$, as indicated in the figure. In the first case, six bonds are deleted in the renormalized sublattice, while only three disapear in for other case.

Now consider the case of a finite concentration $x$ of impurities or holes. Since an impurity belongs to one of the bipartite sublattices, say $A$, there are two effects. The first is a reduction in the average coordination number and the second is that some bonds are deleted. This coordination effect is estimated as follows. The first term of Eq. (9) can be written as an average term plus a correlation of amplitude-coordination ,

$$\sum_i Z_i \gamma_0^2 |c_i(E)|^2 = \langle Z \rangle \gamma_0^2 + V\gamma_0^2 \sum_i \delta Z_i \delta c_i^2(E) \quad (10)$$

where it was used that $Z_i$ can be written as an average $\langle Z \rangle$ plus a fluctuation part $\delta Z_i$. A similar procedure can be made for $|c_i(E)|^2 \equiv \langle c^2(E) \rangle + \delta c_i^2(E)$. The average coordination number can be obtained by observing that around a given carbon atom, there are four possible configurations: it can be surrounded by one, two and three impurities, or it can be completely surrounded by carbon atoms. For each configuration, there is a different coordination number $Z$, since impurities act as holes. As a result, the coordination number $Z$ has a binomial probability distribution $P(Z) = C_Z^3 x^Z (1-x)^{3-Z}$ where $C_Z^3$ is a combinatorial factor. It follows that $\langle Z \rangle$ is the first moment of the binomial distribution: $\langle Z \rangle = \sum_{Z=0}^{Z=3} Z P(Z) = 3(1-x)$. The contribution of the last term in Eq. (10) leads to the production of impurity states, since it is the correlation between amplitude and self-energy fluctuations. Thus, the system has a mobility edge when this term lowers the energy compared with the energy required for having an extended state with amplitude in frustrated bonds.

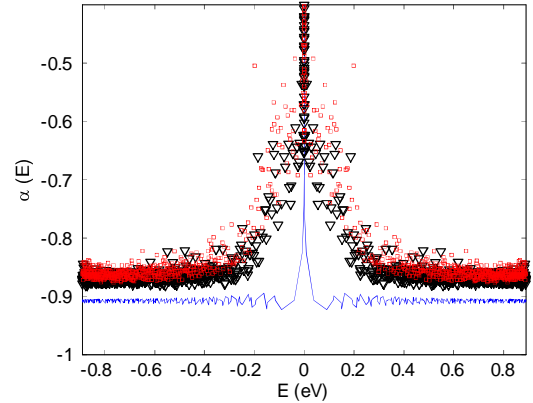

FIG. 3: Logarithm of the inverse participation ratio as a function of the energy for pure graphene (line) compared with the dope case with $x = 0\ 01$ (triangles) and $x = 02$ (squares) around the center of the spectrum of the carbon sub-band, for a lattice with 5184 sites. Observe the rise at the center of the spectrum for the doped case. A band of degenerate states is also observed for pure graphene. The zero corresponds to the Fermi energy.

The other effect is the removal of bonds that changes the second term of Eq. (9). We can estimate this effect as follow. For low concentration of impurities $x \ll 1$, most of them are isolated, since the probability of having two impurities as neighbours goes as $x^2$. Thus, we will consider that impurities are isolated. Two situations are possible. Either an impurity belongs to the renormalized sublattice, or it can remain as shown in Fig. 2. For each impurity site that is renormalized, 3 bonds are lost. In the other case, 6 bonds are lost for each impurity. Since they are randomly distributed in sublattices $A$ and $B$, the concentration of impurities is $x$ on each sublattice. As a result, the number of missing bonds is $(6+3)xN$, from a previous total of $3N$. Using this count in Eq. (9), and assuming no self-energy amplitude correlation Eq. (10) for an antibonding trial state, we obtain the approximate position of the mobility edge $(E_d)$,

$$(E_d - E_0)^2 \approx 3\gamma_0^2(1-x) - \gamma_0^2(3-9x) = 6\gamma_0^2 x,$$

which leads to a symmetric mobility edge separated an energy $\Delta$ from the center of the band,

$$\Delta \approx \pm\sqrt{6x}\gamma_0. \qquad (11)$$

As a check of these ideas, in Fig. 3 we present the normalized logarithm of the inverse participation ratio,

$$\alpha(E) = \frac{\log IP(E)}{\log N}$$

where $IP(E)$ is the inverse participation ratio, defined as $IP(E) = \sum_{i=1}^{N} \| c_i(E) \|^4$, which is a well-known measure of localization. For extended states, $\alpha(E) \approx -1$, while it tends to be bigger values when localization is present. Fig. 3 shows a comparison between pure graphene case and the doped cases, for a tigth-binding simulation using an average of 10 lattices with $N = 5184$ sites. It is worthwhile mentioning that a band of degenerated states appears in the center of pure graphene, which has not been reported previously by other workers. They are a consequence of the local topology of the lattice, as also happens in the square [10] and Penrose lattices, and are due to a decoupling of the $A$ and $B$ sublattice at the center of the spectrum. Fig. 3 shows that the $IP(E)$ is in general bigger for the doped case, but at the center of the spectrum there is a clear rise in its value, indicating a greater degree of localization. In Fig. 4 we compare Eq. (11) with the numerical value of $\Delta$ obtained from the localization plot, which shows a good agreement with the predicted value.

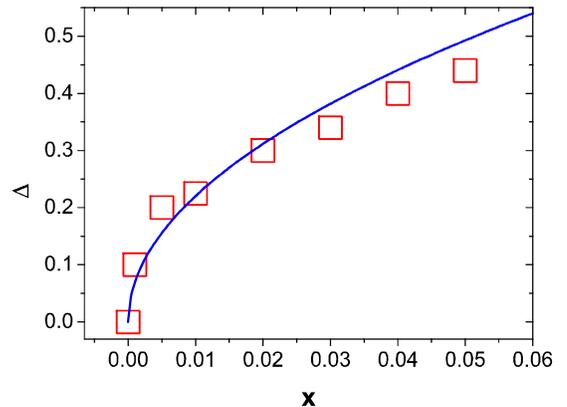

FIG. 4: Theoretical value of the mobility edge predicted from Eq. (11), indicated with a solid line, and the value obtained from a direct diagonalization of the Hamiltonian (squares). The numerical results were obtained from an average of 10 lattices with $N = 5184$ sites

The value of $\gamma_0$ is around [9] $0.9eV$ or $\gamma_0 = 20Kcal/mol$. For a 1% doping, the size of the whole localized region is around $2\Delta \approx 0.44eV$. Since light absorbed when the band-gap energy is in the limit of the visible spectrum $1.77eV$ $700nm$), the localized region in doped graphene can be considered as similar in size as the energy gap in narrow-band- gap semiconductors.

In conclusion, we have shown that doped graphite in the split band regimen presents a mobility edge at the center of the spectrum, an this can be useful for many devices since the position of the mobility edge can be controlled by doping.

**Acknowledgments.** I would like to thank M. and H. Terrones for useful suggestions. This work was supported by DGAPA UNAM project IN108502, and CONACyT 48783-F and 50368.